\documentclass[%
 reprint,
 amsmath,amssymb,
 aps,
]{revtex4-1}

\usepackage{graphicx}
\usepackage{dcolumn}
\usepackage{bm}

\begin{document}

\preprint{APS/123-QED}

\title{First-order Phase Transition and Resulted Expansion in Momentum Space}

\author{Feng Li}
\email{fengli@lzu.edu.cn}
\affiliation{School of Physical Science and Technology, Lanzhou University, Lanzhou, Gansu, 073000, China }

\date{\today}

\begin{abstract}
We point out, according to the principle of entropy growth, that the volume occupied by the system in the momentum space should expand during the first-order phase transition. Such an expansion is visualized by the simulation using the transport model based on a NJL-typed Lagrangian, and quantified by several observables, some of which, proposed for the first time in this article, could be the probes of the first-order phase transition either between the quark gluon plasma and the hadronic phase taking place in the heavy-ion collisions or of the other expanding systems.

\end{abstract}

\pacs{Valid PACS appear here}
\maketitle


Exploring the phase structure of matter in the extreme cases is among the key goals of physics. It is predicted by the quantum chromodynamics that at large temperatures or baryon chemical potentials, a novel matter composed of the liberated quarks and gluons comes into being. The so called  quark-gluon plasma (QGP)~\cite{Shuryak:1980tp} once emerged right after big bang, and is re-created in the laboratory by colliding two heavy-ions at large energies.  The transition from QGP to the normal hadronic matter, according to the lattice QCD calculation~\cite{Bernard:2004je,Aok06,Baz12}, is considered as a cross-over transition at small baryon chemical potentials. However, the transition property at large chemical potentials is still a mystery due to the obstacles from both the theoretical and experimental sides. Theoretically, due to the so-called fermion sign problem ~\cite{PhysRevB.41.9301}, it is hard to carry out the lattice calculation at large baryon chemical potentials. Experimentally, the ultra-hot QGP can not be hold stably under the current technology. Therefore, the phase transition does not take place under a controllable environment and cannot be observed directly. According to numerous phenomenology models~\cite{Nam611,Nam612,Asa89,Car10,Bra12,Stephanov:2004wx,Stephanov:2007fk,Fukushima:2008wg,Fukushima:2013rx,Baym:2017whm,Sun:2020bbn}, a first order phase transition might take place at large baryon chemical potentials, and several observables~\cite{Rischke:1995pe,Stoecker:2004qu,Sun:2017xrx,Sun:2018jhg,Sun:2020pjz} are proposed as the indirect signals. For instance, it is recently proposed that the ratio of the light nuclei yields~\cite{Sun:2017xrx,Sun:2018jhg,Sun:2020pjz} should be enhanced due to the clustering during the phase decomposition process.

\begin{figure}
    \includegraphics[width=0.23\textwidth]{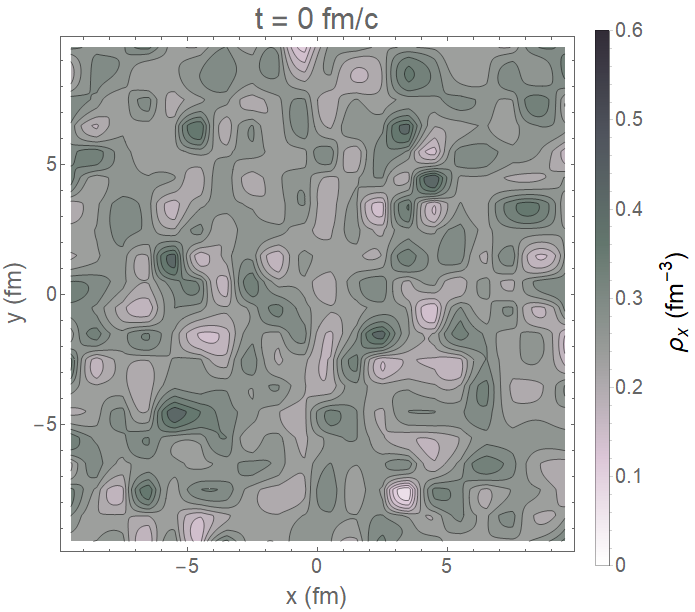}
    \includegraphics[width=0.23\textwidth]{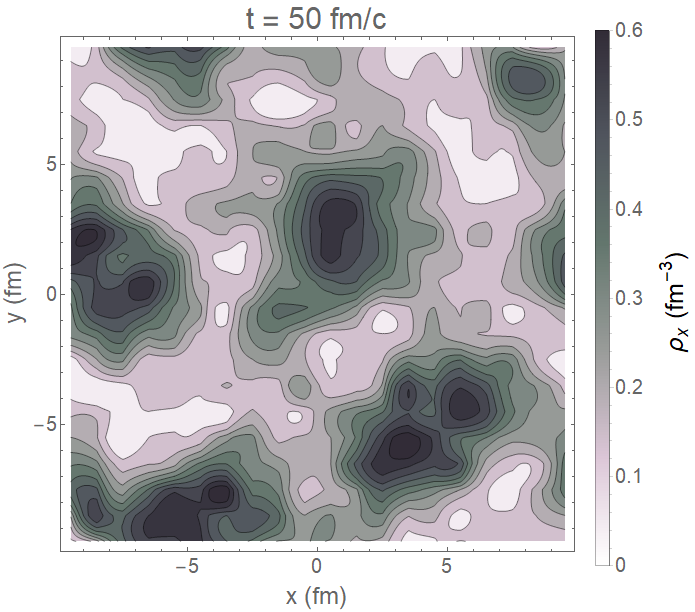}
    \includegraphics[width=0.23\textwidth]{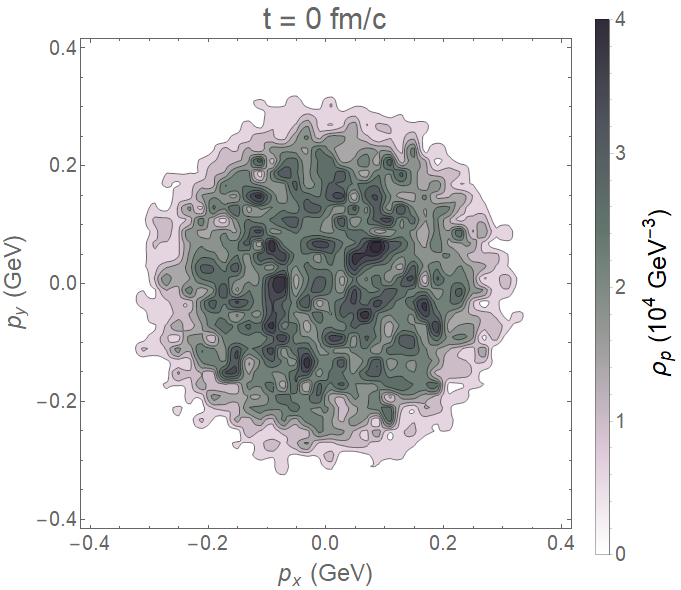}
    \includegraphics[width=0.23\textwidth]{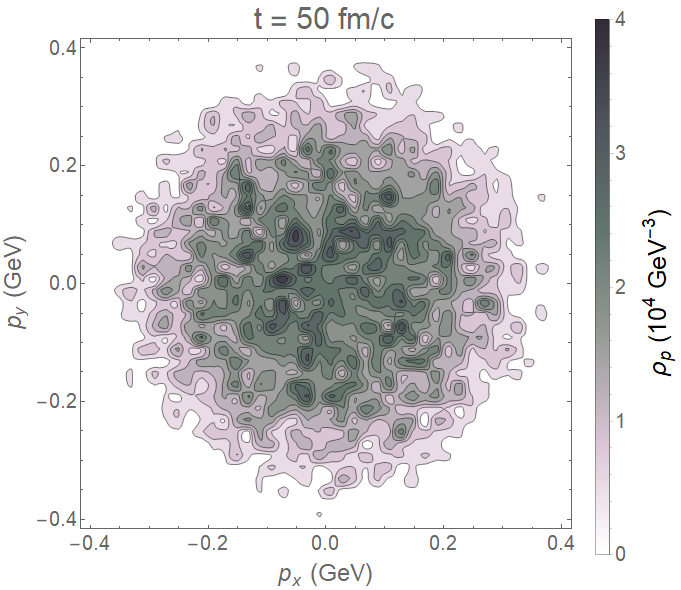}
    \caption{The density distribution of the u-quarks in both the spatial (upper panels) and the momentum (lower panels) space at the beginning (left column) and the end (right column) of the phase transition}
    \label{fig:Visualization}
\end{figure}

But wait, the first-order phase transition or the clustering process is a little counter-intuitive, since it seems against the principle of the increase of entropy, which is usually illustrated by the spread of a drop of ink in a cup of water. It is widely spoken that the structures or in-homogeneity would be eliminated as a result of entropy growth. This is certainly not true. If the entropy within a fixed volume is a convex function of the particle number, it can be easily shown that the system would be of a larger entropy if the particles are not uniformly distributed, and the clusters or the structures thus emerges spontaneously. However, the previous inaccurate opinion does touch the point to some extent, since, as known, the entropy of a system is positively dependent on the volume of the phase space it occupies. The system does occupy a smaller and smaller volume in the spatial space during clustering, which makes the entropy looks decreasing. Therefore, as a compensation, the system must occupies a larger volume in the momentum space, so that the principle of the entropy growth is respected. 

To visualize the expansion in the momentum space resulted by the first order phase transition, we plot in Fig. \ref{fig:Visualization} the u-quark distribution in both the spatial (upper panels) and the momentum (lower panels) space at the beginning (left column) and the end (right column) of the phase transition, respectively. The simulation is carried out by solving the transport equations based on a SU(3) NJL-type Lagrangian~\cite{Bra12}. Readers might refer to Ref. \cite{Li:2016uvu} for the details of the transport model. The 20 fm $\times$ 20 fm $\times$ 20 fm box is initially filled with the uniformly and thermally distributed quarks with the temperature $T=20$ MeV and the net quark density $n_q = 0.5$ fm$^{-3}$. A periodic boundary condition is employed. The system is initially spinodal unstable, and then undergoes phase decomposition. As shown, the dense clusters are generated spontaneously, and meanwhile, the u-quarks do expand in the momentum space. The particles boosted to the large momentum region are most probably located inside the clusters. They gain extra kinetic energies in the cost of losing their potential energies. 

In the following sections, we define several observables to quantify the expansion in the momentum space, some of which might be considered as the new probes of the first-order phase transition. 

\begin{enumerate}

    \item The quadratic mean of the momenta
    
    The quadratic mean of the momenta, defined as
    \begin{equation}
        \sqrt{\overline{p^2}} \equiv \sqrt{\frac{1}{N} \int \frac{d^3\mathbf p d^3 \mathbf x}{(2\pi)^3}f(\mathbf x, \mathbf p) \mathbf p^2} 
    \end{equation}
    with $f(\mathbf x, \mathbf p)$ being the phase space distribution of the u-quarks and 
    \begin{equation}
        N \equiv  \int \frac{d^3\mathbf p d^3 \mathbf x}{(2\pi)^3}f(\mathbf x, \mathbf p)  
    \end{equation}
    being the total u-quark number, is a quantity positively dependent on the volume occupied in the momentum space, and as shown in fig. {\ref{fig:QuadraticMean}}, does grow with time during the phase decomposition process. However, the quadratic mean of the momenta might not be a proper probe of the first-order phase transition, since the enhancement caused by the phase decomposition might be covered by the thermal background, which is not easy to be removed.
    
    \begin{figure}
    \includegraphics[width=0.4\textwidth]{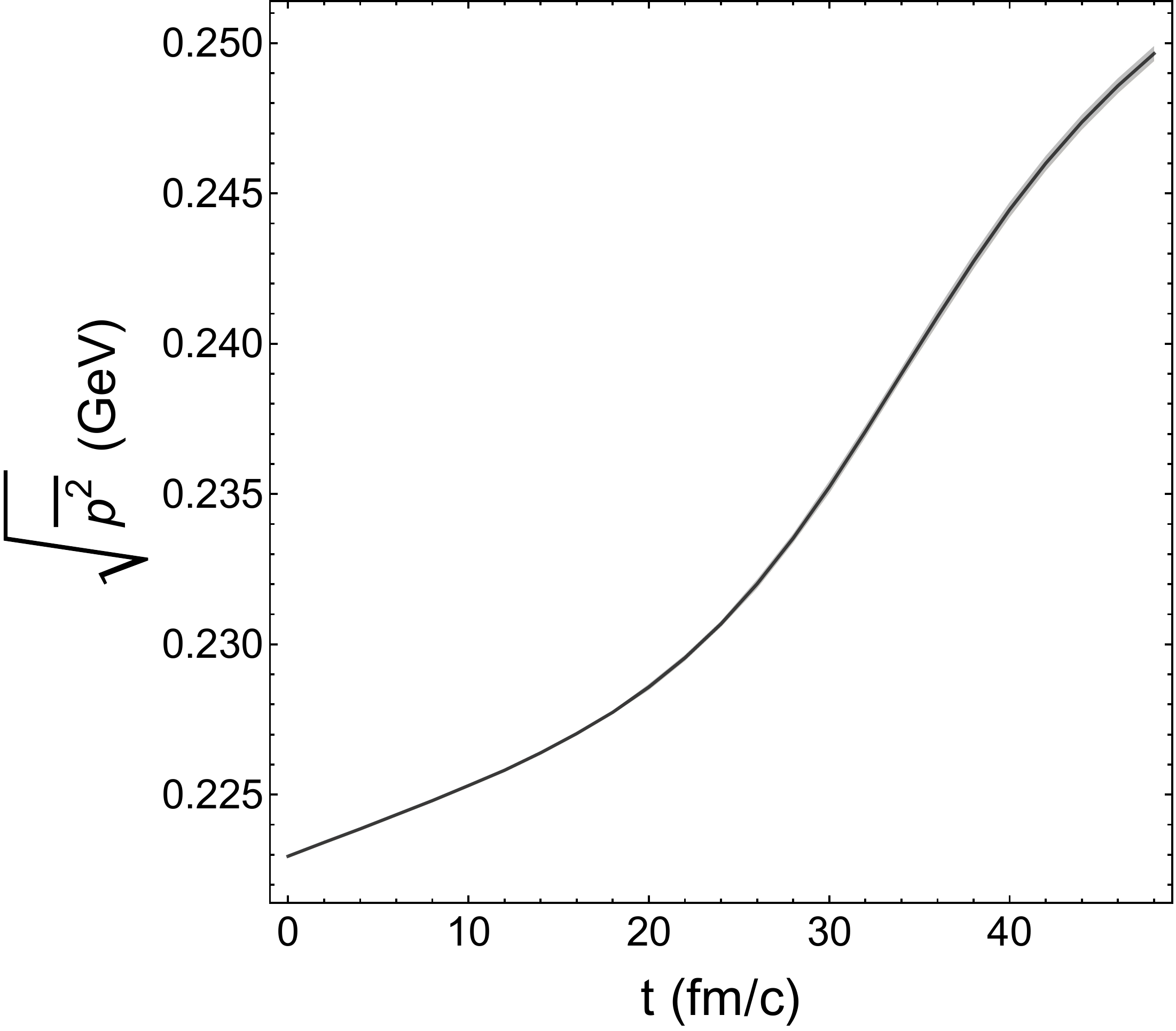}
    \caption{The evolution of the quadratic mean of the u-quark momenta}
    \label{fig:QuadraticMean}
    \end{figure}
    
    \item Average density in the spatial and momentum space
    
    The shrinking in the spatial space due to clustering can be characterized by the enhancement of the average density in the spatial space defined as
    \begin{equation}
        \overline{\rho_x} \equiv \frac 1 N\int d^3\mathbf x \rho(\mathbf x)^2
    \end{equation}
    where 
    \begin{equation}
        \rho(\mathbf x) \equiv \int \frac{d^3\mathbf p}{(2\pi)^3} f(\mathbf x, \mathbf p)
    \end{equation}
    is the local particle density. Similarly, the resulted expansion in the momentum space should lead to the reduction of the average density in the momentum space, which is defined as
    \begin{equation}
        \overline{\rho_p} \equiv \frac 1 N\int \frac{d^3\mathbf p}{(2\pi)^3} \rho(\mathbf p)^2
    \end{equation}
    where 
    \begin{equation}
        \rho(\mathbf p) \equiv \int d^3\mathbf x f(\mathbf x, \mathbf p)
    \end{equation}
    is the density in the momentum space. The time evolution, during the phase decomposition process, of the average densities of the u-quarks defined in both the coordinate and the momentum spaces are plotted in fig. \ref{fig:AvgDensity} by the solid and the dashed lines respectively. As expected, they do show a growing and a declining behavior respectively, indicating the shrinking in the coordinate space and the expansion in the momentum space. 
    
    \begin{figure}
    \includegraphics[width=0.4\textwidth]{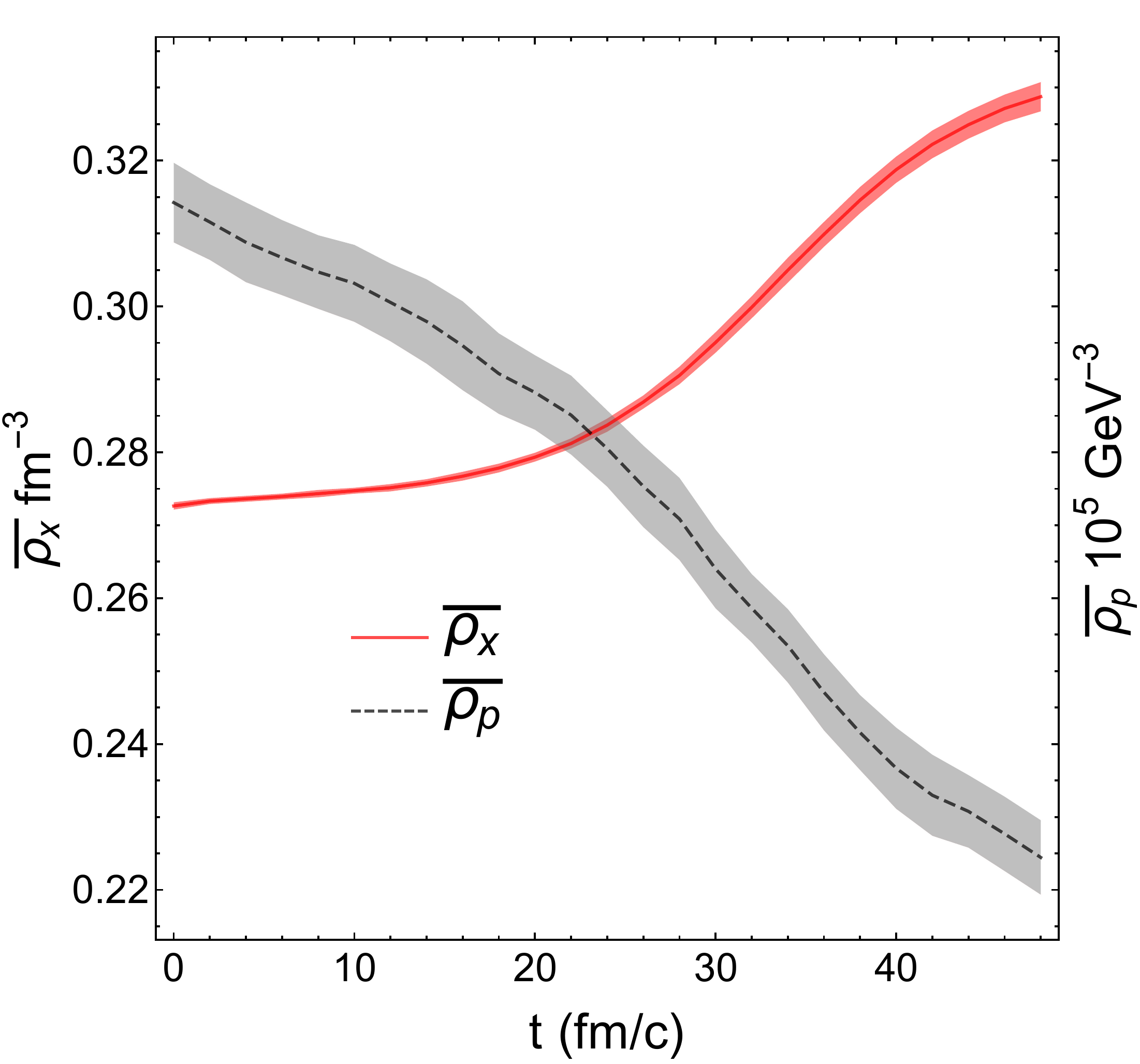}
    \caption{The evolution of the average density of the u-quarks 
    in both the spatial and the momentum spaces.}
    \label{fig:AvgDensity}
    \end{figure}
    
    The average density of the u-quarks, or any particle species, in the momentum space could be employed as the probe of the first order phase transition, if its thermal background can be removed. It can be easily shown that, if the phase space distribution is replaced with the Fermi-Dirac function, i.e.,
    \begin{equation}
        f(\mathbf x, \mathbf p) = \left(e^{\frac{E_p-\mu}{T}}+1\right)^{-1},
    \end{equation}
    the thermal expectation value of $\overline{\rho_p}$ is
    \begin{equation}
        \langle \overline{\rho_p} \rangle_{th} = V_\mathrm{fireball} \left(1-\frac{\langle\delta N^2\rangle_{th}}{\langle N\rangle_{th}}\right)
    \end{equation}
    where $\delta N$ is the event by event fluctuation of the particle multiplicity, and $V_\mathrm{fireball}$ is the volume of the fireball at the chemical freeze-out, which can be evaluated by the ratio of the total baryon number and the baryon density at hadronization (or the ratio of the total particle energy and the energy density at hadronization). Without the first-order phase transition, partons are thermally distributed during the partonic phase in the heavy-ion collisions, therefore the ratio $\overline{\rho_p}/\langle \overline{\rho_p}\rangle_{th}$ should always be one. Due to the suppression of $\overline{\rho_p}$ aroused by the first-order phase transition, we expect the appearance of a dip on the  $\sqrt{s_{NN}}-\overline{\rho_p}/\langle \overline{\rho_p}\rangle_{th}$ curve as shown in fig.\ref{fig:AvgRhoPBES}, if the system really goes through the first-order phase decomposition at certain colliding energies, and can therefore be measured in the beam energy scan (BES)~\cite{Bzdak:2019pkr} project. 
    
    \begin{figure}
        \centering
        \includegraphics[width=0.4\textwidth]{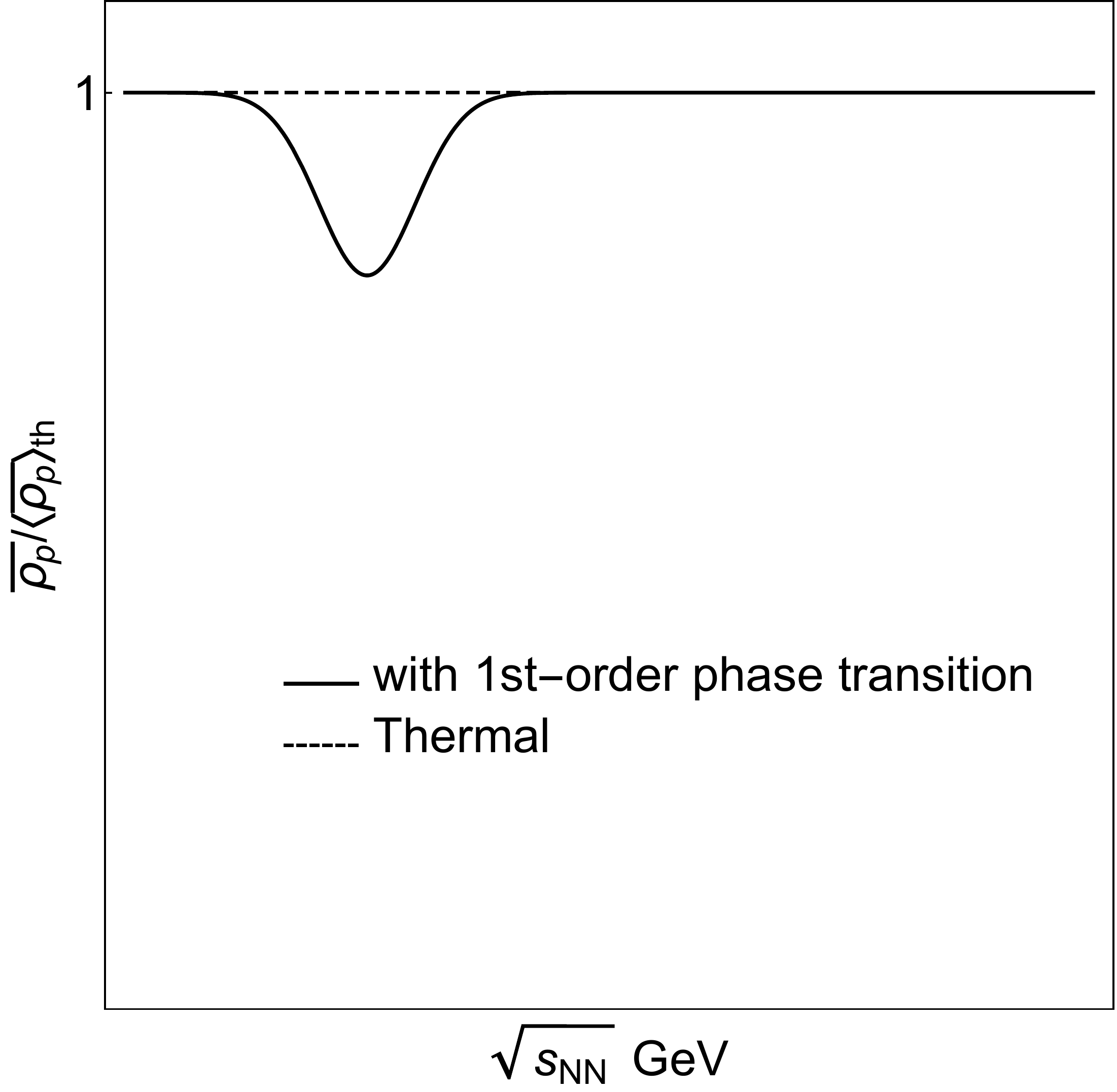}
        \caption{$\bar \rho_p$ should be below its thermal expectation value if the system undergoes a fist-order phase transition at certain colliding energies, and thus leads to a dip on the $\overline{\rho_p}/\langle \rho_p\rangle_{th} - \sqrt{s_{NN}}$ curve.}
        \label{fig:AvgRhoPBES}
    \end{figure}
    
    However, to employ $\overline{\rho_p}/\langle \overline{\rho_p}\rangle_{th}$ as the probe of the first-order phase transition, one still need to assign an assumed value to either the baryon number density or the energy density at hadronization for evaluating $\langle \overline{\rho_p}\rangle_{th}$. To avoid such an ambiguity, we then investigate the next observable.
    
    \item Scaled density moments in the spatial and momentum space
    
    It is illustrated in Ref. ~\cite{Steinheimer:2012gc} that the $n^\mathrm{th}$-order scaled density moments, defined as
    \begin{equation}
        y_n \equiv \frac 1 N \frac{\int d^3 \mathbf x \rho(\mathbf x)^{n+1}}{\overline{\rho_x}^n},
    \end{equation}
    quantifies to what extent the particle distribution deviates from a uniform one. $y_n = 1$, if the distribution is uniform, while $y_n > 1$, if the distribution is in-homogeneous. Therefore, $y_n$ is enhanced due to clustering during the phase decomposition process.
    
    Such a definition can be extended to the momentum space as
    \begin{equation}
        \psi_n \equiv \frac 1 N \frac{\int \frac{d^3 \mathbf p}{(2\pi)^3} \rho(\mathbf p)^{n+1}}{\overline{\rho_p}^n},
    \end{equation}
    measuring the deviation from the uniform distribution in the momentum space. At large baryon chemical potentials, the expansion or diffusion in the momentum space aroused by the first-order phase transition can be considered as a smearing of the Fermi-Dirac function, and therefore enhances $\psi_n$ as well. Think of the extreme case with a vanishing temperature. The particles are uniformly distributed inside the Fermi-sphere in the momentum space, and $\psi_n$ is thus 1. Shifting any particle out of the Fermi-sphere for occupying a larger volume in the momentum space would modify such a uniform configuration and make $\psi_n$ greater than one.
    
    The enhancement of both the $y_n$ and $\psi_n$ during the first order phase transition are shown via the transport simulation and plotted in fig. \ref{fig:Moments} by the solid and the dashed lines for $n=2$ and $n=3$ , respectively.
    
    \begin{figure}
    \includegraphics[width=0.4\textwidth]{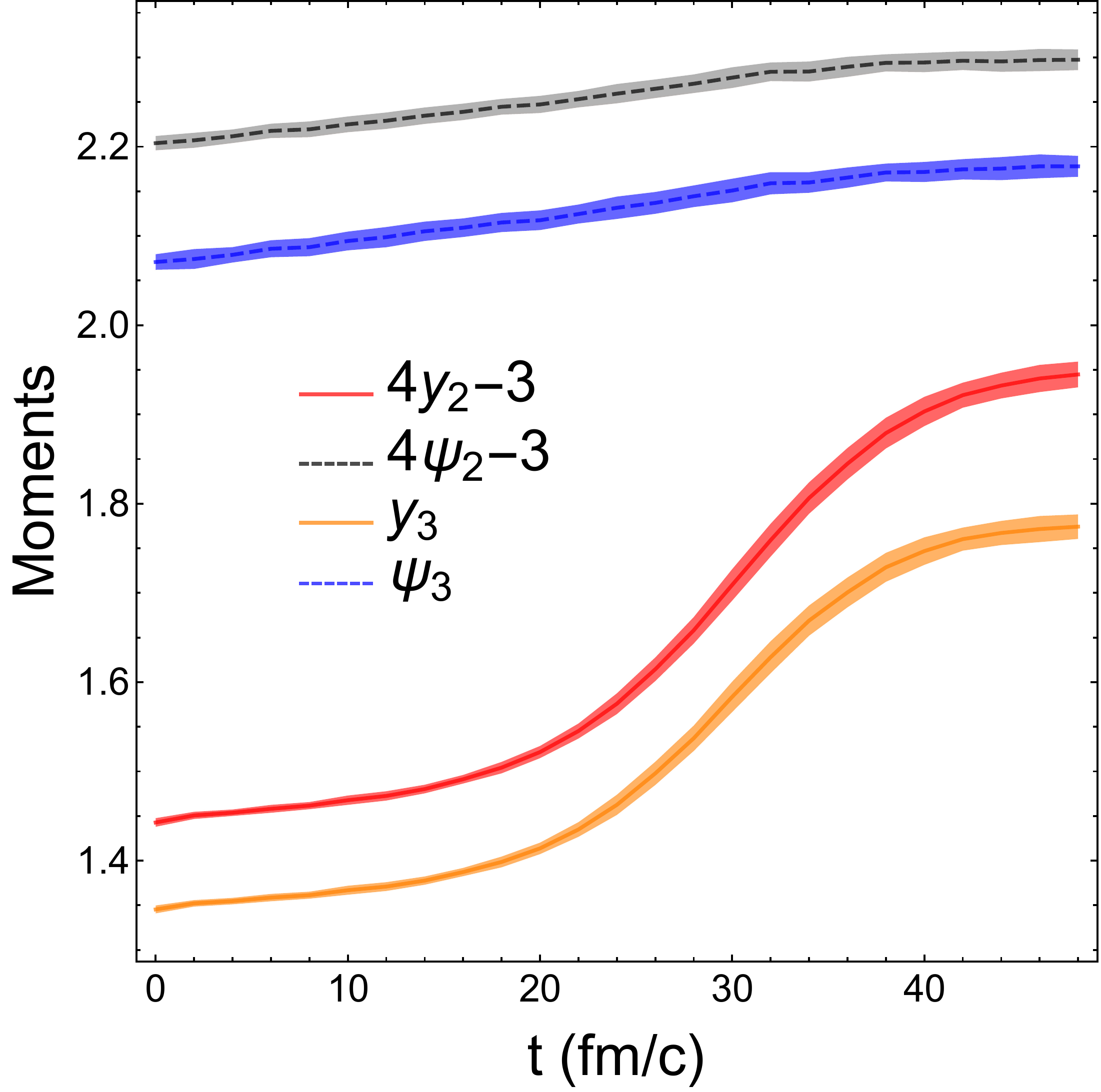}
    \caption{The time evolution of the second and third order density moments in both the spatial and the momentum spaces.}
    \label{fig:Moments}
    \end{figure}
    
    The thermal expectation value of $\psi_2$ and $\psi_3$ are
    \begin{eqnarray}
    \label{eq:psi2}
    \langle \psi_2\rangle_{th} &=& \frac{\langle N \rangle_{th}}{2} \frac{\langle\delta N^3 - 3 \delta N^2 + 2 N\rangle_{th}}{\langle N - \delta N^2 \rangle_{th}^2}, \\
    \label{eq:psi3}
    \langle \psi_3\rangle_{th} &=& \langle N \rangle_{th}^2 \frac{\langle -\frac 1 6 \delta N^4 + \delta N^3 -\frac {11} 6 \delta N^2 + N \rangle_{th}}{\langle N - \delta N^2\rangle_{th}^3}\nonumber\\
    && + \frac{\langle N \rangle_{th}^2}{2} \frac{ \langle \delta N^2 \rangle_{th}^2}{\langle N - \delta N^2\rangle_{th}^3}.
    \end{eqnarray}
    Notice that the right hand sides of Eq.(\ref{eq:psi2}) and Eq.(\ref{eq:psi3}) are composed of the quantities which are directly measurable. Therefore, the ratios $\psi_n / \langle \psi_n \rangle_{th}$ for $n=2$ and $3$ can be employed as the signals for the first order phase transition. Their values should exceed one due to the enhancement effect in the presence of the first order phase transition. Otherwise, the ratios should always be equal to one. We therefore expect the appearance of a peak on the $\sqrt{s_{NN}}-\psi_n/\langle\psi_n\rangle_{th}$ curve as shown in fig. \ref{fig:MomentBES}, if the systems undergoes the first order phase decomposition at certain colliding energies and can therefore be checked in the BES experiments.
    \begin{figure}
        \centering
        \includegraphics[width=0.4\textwidth]{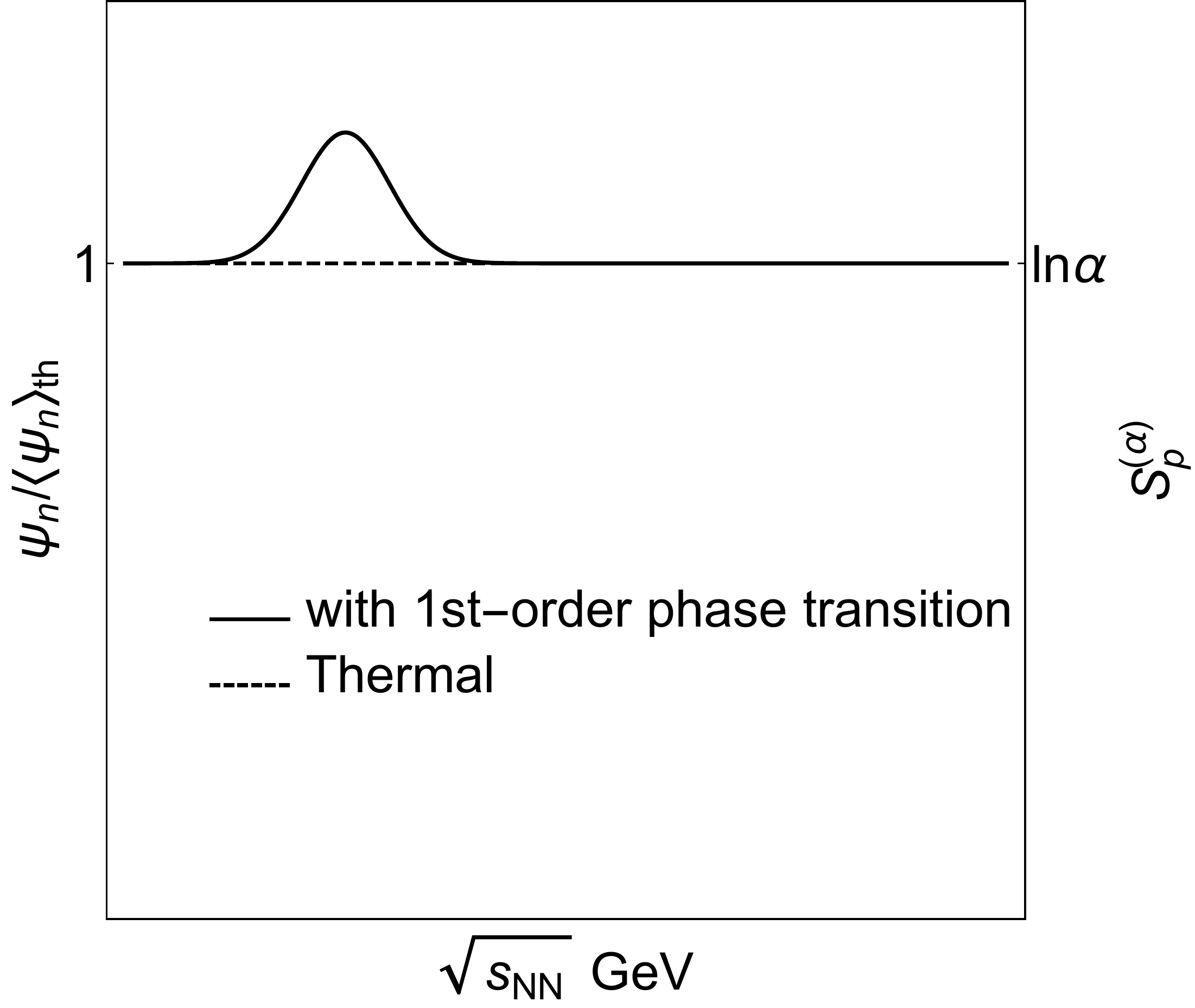}
        \caption{$\psi_n$ should be above its thermal expectation value, and $S^{(\alpha)}_p$ should exceed $\ln \alpha$, if the system undergoes a first-order phase transition at certain colliding energies, and thus leads to a peak on both the $\sqrt{s_{NN}}-\psi_n/\langle\psi_n\rangle_{th}$ and $\sqrt{s_{NN}}-S^{(\alpha)}_p$ curves.}
        \label{fig:MomentBES}
    \end{figure}
    
    \item Spatial and momentum entropies
    
    The last quantities of interest are the spatial and momentum entropies defined as 
    \begin{eqnarray}
    S_x &=& -\int d^3 \mathbf x \frac{\rho(\mathbf x)}{N} \ln \frac{\rho(\mathbf x) v_x}{N}, \\
    S_p &=& -\int \frac{d^3 \mathbf p}{(2\pi)^3} \frac{\rho(\mathbf p)}{N} \ln \frac{\rho(\mathbf p) v_p}{N},
    \end{eqnarray}
    where the constants $v_x$ and $v_p$ are the unit volumes in the spatial and the momentum spaces. Since $P_x = v_x \rho(\mathbf x) / N$ is the probability for a particle being located in the vicinity of volume $v_x$ around $\mathbf x$, $S_x = -\sum_x P_x \ln P_x$ is exactly the Shannon entropy describing the complexity of the spatial configuration. Similarly, $S_p$ is the Shannon entropy describing the complexity of the configuration in the momentum space. Changing $v_a \to \alpha v_a$, with $a=x$ or $p$ representing either the spatial or the momentum components respectively, shifts the value of the entropies only by a constant equal to $-\ln \alpha$. Therefore, the evolution trends of the entropies are independent of the choice of the unit volumes.  
    
    The evolution, during the phase decomposition, of both $S_x$ and $S_p$ with $v_x$ being $(0.5 \mathrm{fm})^3$ and $v_p$ being $(12\mathrm{MeV})^3$ are plotted in fig. \ref{fig:AvgEntropies} by the solid and dashed lines respectively. As expected, $S_x$ decreases with time, 
    illustrating the origin of the public bias, i.e., given only the spatial coordinates, the entropy seems declining during the clustering process. While, as a compensation, the entropy calculated using only the particle momenta grows. 
    \begin{figure}
    \includegraphics[width=0.4\textwidth]{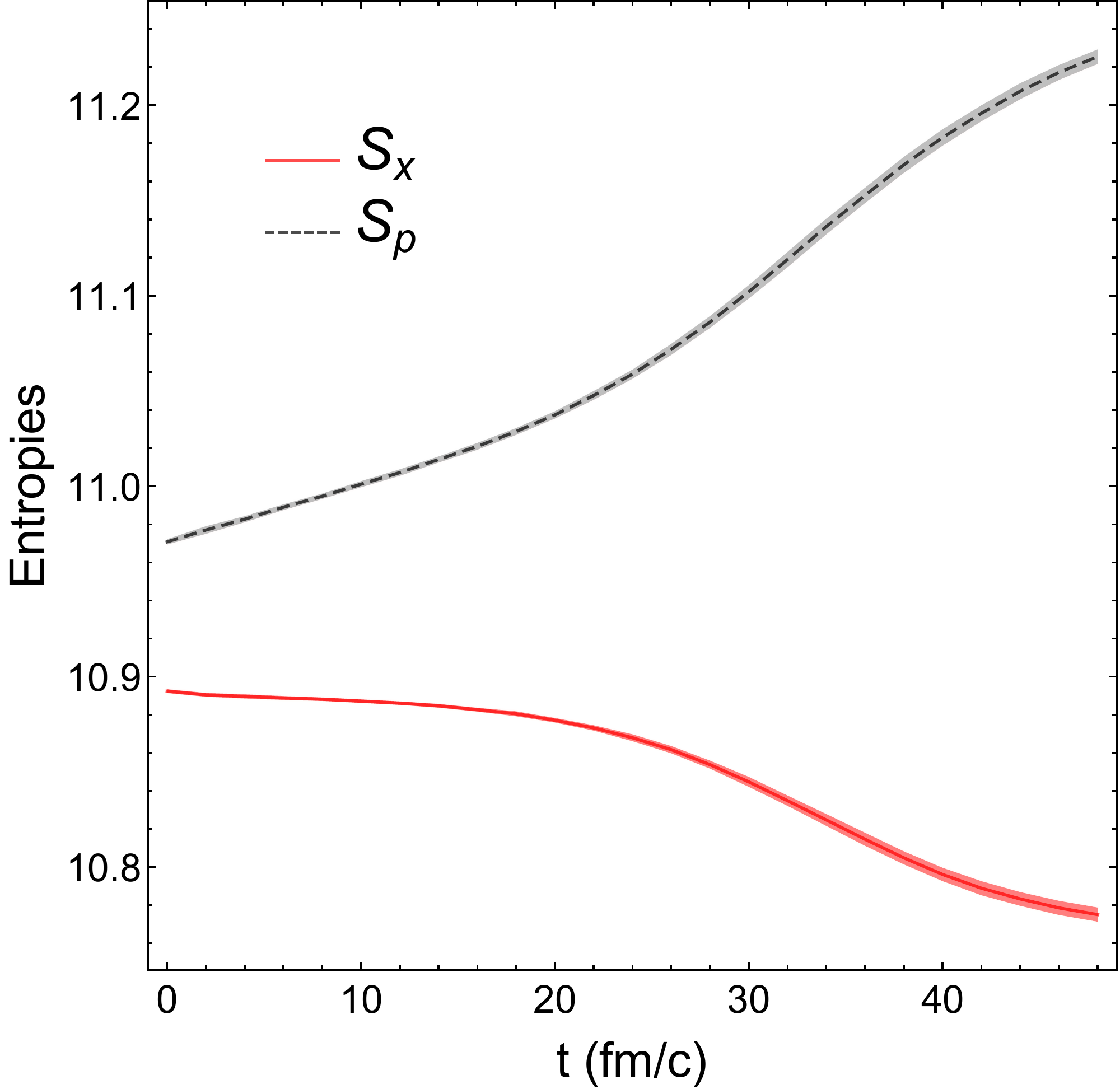}
    \caption{The evolution of both the spatial and the momentum entropies with $v_x$ being $(0.5 \mathrm{fm})^3$ and $v_p$ being $(12\mathrm{MeV})^3$.}
    \label{fig:AvgEntropies}
    \end{figure}
    
    The enhancement of $S_p$ might be considered as a signal of the first-order phase transition as well. The thermal background of $S_p$ can be removed by adjusting the value of the unit volume in the momentum space $v_p$ according to the temperatures and. If
    \begin{equation}
        v_p = v^{(\alpha)}_p \equiv \alpha^{-1}\frac{\langle N \rangle_{th}}{V_\mathrm{fireball}} \exp \left(\frac{\langle S \rangle_{th}}{\langle N \rangle_{th}}\right)
    \end{equation}
    where $\langle S \rangle_{th} / \langle N \rangle_{th}$ is the thermal expectation value of the entropy per particle at hadronization, and $\alpha$ is a constant coefficient, the thermal expectation value of $\langle S_p\rangle_{th} =  \ln \alpha$ is a constant. We plot $v^{(\alpha)_p}$ in fig. \ref{fig:ThermalVp} as the function of the temperature and the baryon chemical potential (upper panel) and thus the colliding energies of the BES experiments (lower panel) by assuming the the quark mass equal to $0.1$ GeV at hadronization. The temperatures and the chemical potentials, at chemical freeze-out, of the fireballs created from the collisions in the BES experiments taken from the Ref. ~\cite{PhysRevC.96.044904}. So,
    \begin{eqnarray}
    S^{(\alpha)}_p \equiv -\int \frac{d^3 \mathbf p}{(2\pi)^3} \frac{\rho(\mathbf p)}{N} \ln \frac{\rho(\mathbf p) v^{(\alpha)}_p}{N} > \ln \alpha
    \end{eqnarray}
    can be the signal indicating the system once undergoes the first-order phase transition, and thus, as shown in fig. \ref{fig:MomentBES}, leads to a peak on the $\sqrt{s_{NN}} - S^{(\alpha)}_p$ curve.
    \begin{figure}
        \centering
        \includegraphics[width=0.4\textwidth]{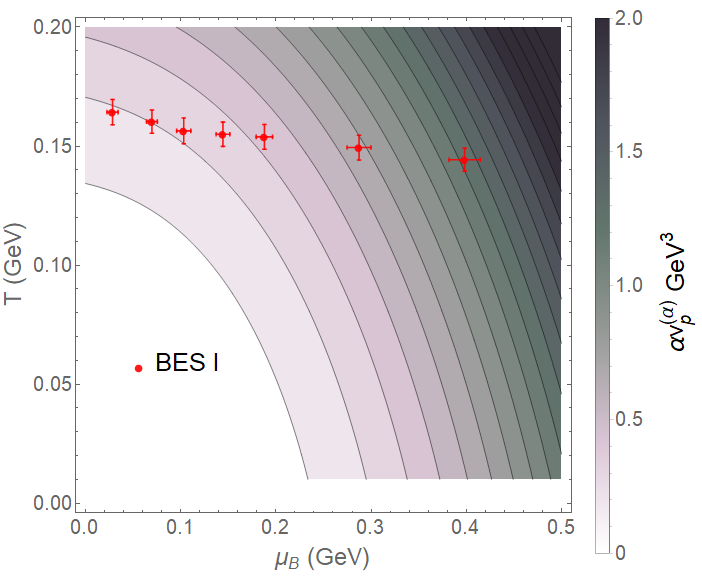}
        \includegraphics[width=0.35\textwidth]{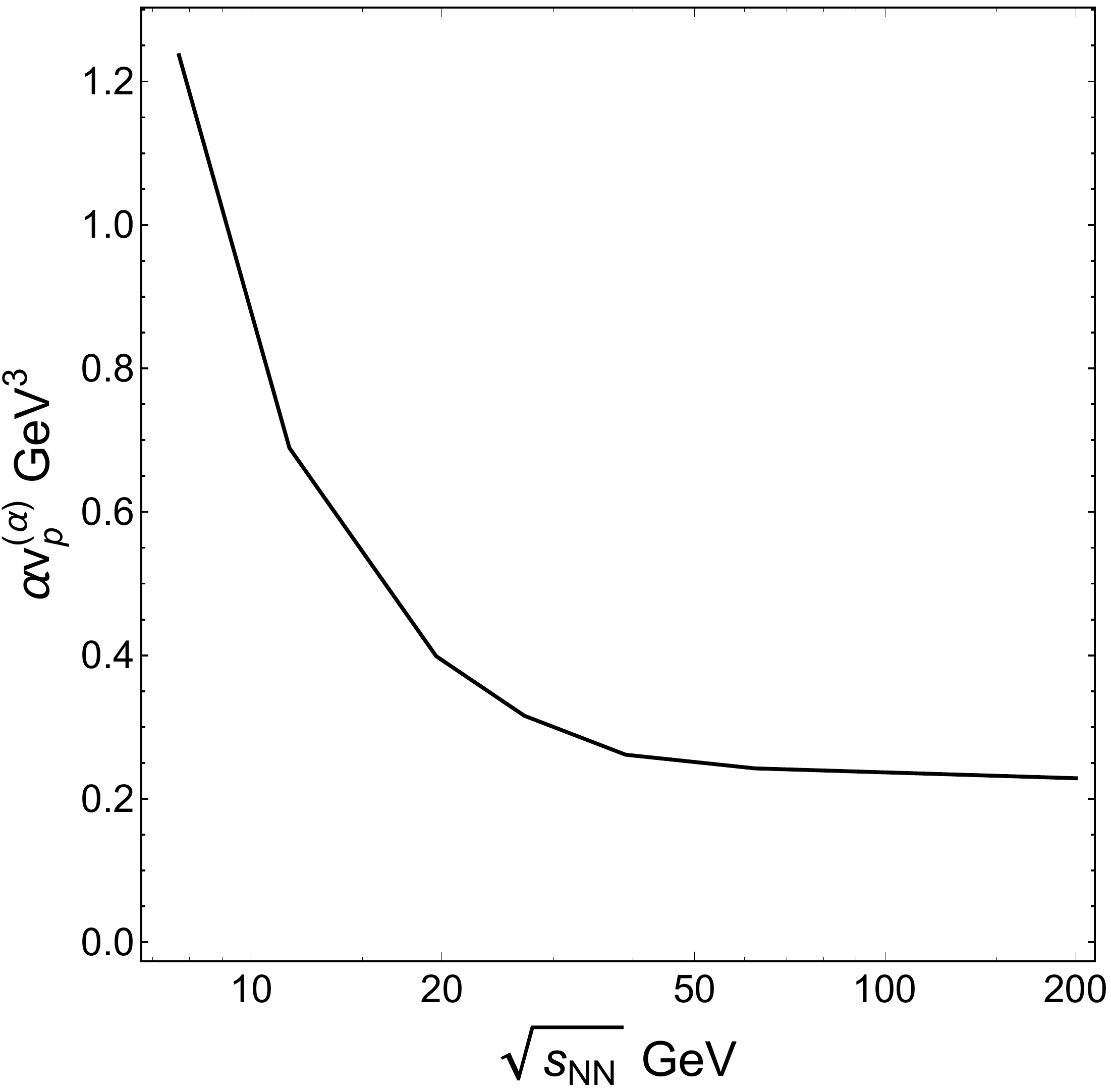}
        \caption{The value of the unit volumes in the momentum space $v^{(\alpha)}_p$, which is chosen so that the thermal expectation value of the momentum entropy $\langle S_p \rangle_{th}$ is independent of $T$ and $\mu_B$, as the function of the temperature and the baryon chemical potential (upper panel) and thus the colliding energies of the BES experiments (lower panel).}
        \label{fig:ThermalVp}
    \end{figure}
\end{enumerate}

In summary, we point out, according to the principle of entropy growth, that the particles should expand in the momentum space during the first-order phase transition. Such an expansion is visualized via the simulations using the transport model based on a NJL-typed Lagrangian. Several quantities, including the quadratic mean of the momenta $\sqrt{\overline{p^2}}$, the average density in the momentum space $\rho_p$, the scaled density moments in the momentum space $\psi_n$, and the momentum entropy $S_p$ are proposed to quantify such an expansion, and the latter three can be employed as the signals of the first-order phase transition either between the QGP and the hadronic phase in the heavy-ion collisions, or of the other expanding systems such as the nuclear matter undergoes the liquid-gas transition. 

More realistic simulations mimicking the systems created in the BES experiments are carried out to check whether these signals survive the hadronic re-scattering.  

\section*{Acknowledgement}

This work is supported by the Fundamental Research Funds for the Central Universities under Grants No. lzujbky-2021-sp24

\newpage
\bibliography{references}

\begin{thebibliography}{25}%
\makeatletter
\providecommand \@ifxundefined [1]{%
 \@ifx{#1\undefined}
}%
\providecommand \@ifnum [1]{%
 \ifnum #1\expandafter \@firstoftwo
 \else \expandafter \@secondoftwo
 \fi
}%
\providecommand \@ifx [1]{%
 \ifx #1\expandafter \@firstoftwo
 \else \expandafter \@secondoftwo
 \fi
}%
\providecommand \natexlab [1]{#1}%
\providecommand \enquote  [1]{``#1''}%
\providecommand \bibnamefont  [1]{#1}%
\providecommand \bibfnamefont [1]{#1}%
\providecommand \citenamefont [1]{#1}%
\providecommand \href@noop [0]{\@secondoftwo}%
\providecommand \href [0]{\begingroup \@sanitize@url \@href}%
\providecommand \@href[1]{\@@startlink{#1}\@@href}%
\providecommand \@@href[1]{\endgroup#1\@@endlink}%
\providecommand \@sanitize@url [0]{\catcode `\\12\catcode `\$12\catcode
  `\&12\catcode `\#12\catcode `\^12\catcode `\_12\catcode `\%12\relax}%
\providecommand \@@startlink[1]{}%
\providecommand \@@endlink[0]{}%
\providecommand \url  [0]{\begingroup\@sanitize@url \@url }%
\providecommand \@url [1]{\endgroup\@href {#1}{\urlprefix }}%
\providecommand \urlprefix  [0]{URL }%
\providecommand \Eprint [0]{\href }%
\providecommand \doibase [0]{http://dx.doi.org/}%
\providecommand \selectlanguage [0]{\@gobble}%
\providecommand \bibinfo  [0]{\@secondoftwo}%
\providecommand \bibfield  [0]{\@secondoftwo}%
\providecommand \translation [1]{[#1]}%
\providecommand \BibitemOpen [0]{}%
\providecommand \bibitemStop [0]{}%
\providecommand \bibitemNoStop [0]{.\EOS\space}%
\providecommand \EOS [0]{\spacefactor3000\relax}%
\providecommand \BibitemShut  [1]{\csname bibitem#1\endcsname}%
\let\auto@bib@innerbib\@empty
\bibitem [{\citenamefont {Shuryak}(1980)}]{Shuryak:1980tp}%
  \BibitemOpen
  \bibfield  {author} {\bibinfo {author} {\bibfnamefont {E.~V.}\ \bibnamefont
  {Shuryak}},\ }\href {\doibase 10.1016/0370-1573(80)90105-2} {\bibfield
  {journal} {\bibinfo  {journal} {Phys. Rept.}\ }\textbf {\bibinfo {volume}
  {61}},\ \bibinfo {pages} {71} (\bibinfo {year} {1980})}\BibitemShut {NoStop}%
\bibitem [{\citenamefont {Bernard}\ \emph {et~al.}(2005)\citenamefont
  {Bernard}, \citenamefont {Burch}, \citenamefont {Gregory}, \citenamefont
  {Toussaint}, \citenamefont {DeTar}, \citenamefont {Osborn}, \citenamefont
  {Gottlieb}, \citenamefont {Heller},\ and\ \citenamefont
  {Sugar}}]{Bernard:2004je}%
  \BibitemOpen
  \bibfield  {author} {\bibinfo {author} {\bibfnamefont {C.}~\bibnamefont
  {Bernard}}, \bibinfo {author} {\bibfnamefont {T.}~\bibnamefont {Burch}},
  \bibinfo {author} {\bibfnamefont {E.~B.}\ \bibnamefont {Gregory}}, \bibinfo
  {author} {\bibfnamefont {D.}~\bibnamefont {Toussaint}}, \bibinfo {author}
  {\bibfnamefont {C.~E.}\ \bibnamefont {DeTar}}, \bibinfo {author}
  {\bibfnamefont {J.}~\bibnamefont {Osborn}}, \bibinfo {author} {\bibfnamefont
  {S.}~\bibnamefont {Gottlieb}}, \bibinfo {author} {\bibfnamefont {U.~M.}\
  \bibnamefont {Heller}}, \ and\ \bibinfo {author} {\bibfnamefont
  {R.}~\bibnamefont {Sugar}} (\bibinfo {collaboration} {MILC}),\ }\href
  {\doibase 10.1103/PhysRevD.71.034504} {\bibfield  {journal} {\bibinfo
  {journal} {Phys. Rev.}\ }\textbf {\bibinfo {volume} {D71}},\ \bibinfo {pages}
  {034504} (\bibinfo {year} {2005})},\ \Eprint
  {http://arxiv.org/abs/hep-lat/0405029} {arXiv:hep-lat/0405029 [hep-lat]}
  \BibitemShut {NoStop}%
\bibitem [{\citenamefont {Aoki}\ \emph {et~al.}(2006)\citenamefont {Aoki},
  \citenamefont {Endrodi}, \citenamefont {Fodor}, \citenamefont {Katz},\ and\
  \citenamefont {Szabo}}]{Aok06}%
  \BibitemOpen
  \bibfield  {author} {\bibinfo {author} {\bibfnamefont {Y.}~\bibnamefont
  {Aoki}}, \bibinfo {author} {\bibfnamefont {G.}~\bibnamefont {Endrodi}},
  \bibinfo {author} {\bibfnamefont {Z.}~\bibnamefont {Fodor}}, \bibinfo
  {author} {\bibfnamefont {S.~D.}\ \bibnamefont {Katz}}, \ and\ \bibinfo
  {author} {\bibfnamefont {K.~K.}\ \bibnamefont {Szabo}},\ }\href {\doibase
  10.1038/nature05120} {\bibfield  {journal} {\bibinfo  {journal} {Nature}\
  }\textbf {\bibinfo {volume} {443}},\ \bibinfo {pages} {675} (\bibinfo {year}
  {2006})},\ \Eprint {http://arxiv.org/abs/hep-lat/0611014}
  {arXiv:hep-lat/0611014 [hep-lat]} \BibitemShut {NoStop}%
\bibitem [{\citenamefont {Bazavov}\ \emph {et~al.}(2012)\citenamefont {Bazavov}
  \emph {et~al.}}]{Baz12}%
  \BibitemOpen
  \bibfield  {author} {\bibinfo {author} {\bibfnamefont {A.}~\bibnamefont
  {Bazavov}} \emph {et~al.},\ }\href {\doibase 10.1103/PhysRevD.85.054503}
  {\bibfield  {journal} {\bibinfo  {journal} {Phys. Rev.}\ }\textbf {\bibinfo
  {volume} {D85}},\ \bibinfo {pages} {054503} (\bibinfo {year} {2012})},\
  \Eprint {http://arxiv.org/abs/1111.1710} {arXiv:1111.1710 [hep-lat]}
  \BibitemShut {NoStop}%
\bibitem [{\citenamefont {Loh}\ \emph {et~al.}(1990)\citenamefont {Loh},
  \citenamefont {Gubernatis}, \citenamefont {Scalettar}, \citenamefont {White},
  \citenamefont {Scalapino},\ and\ \citenamefont {Sugar}}]{PhysRevB.41.9301}%
  \BibitemOpen
  \bibfield  {author} {\bibinfo {author} {\bibfnamefont {E.~Y.}\ \bibnamefont
  {Loh}}, \bibinfo {author} {\bibfnamefont {J.~E.}\ \bibnamefont {Gubernatis}},
  \bibinfo {author} {\bibfnamefont {R.~T.}\ \bibnamefont {Scalettar}}, \bibinfo
  {author} {\bibfnamefont {S.~R.}\ \bibnamefont {White}}, \bibinfo {author}
  {\bibfnamefont {D.~J.}\ \bibnamefont {Scalapino}}, \ and\ \bibinfo {author}
  {\bibfnamefont {R.~L.}\ \bibnamefont {Sugar}},\ }\href {\doibase
  10.1103/PhysRevB.41.9301} {\bibfield  {journal} {\bibinfo  {journal} {Phys.
  Rev. B}\ }\textbf {\bibinfo {volume} {41}},\ \bibinfo {pages} {9301}
  (\bibinfo {year} {1990})}\BibitemShut {NoStop}%
\bibitem [{\citenamefont {Nambu}\ and\ \citenamefont
  {Jona-Lasinio}(1961{\natexlab{a}})}]{Nam611}%
  \BibitemOpen
  \bibfield  {author} {\bibinfo {author} {\bibfnamefont {Y.}~\bibnamefont
  {Nambu}}\ and\ \bibinfo {author} {\bibfnamefont {G.}~\bibnamefont
  {Jona-Lasinio}},\ }\href {\doibase 10.1103/PhysRev.122.345} {\bibfield
  {journal} {\bibinfo  {journal} {Phys. Rev.}\ }\textbf {\bibinfo {volume}
  {122}},\ \bibinfo {pages} {345} (\bibinfo {year}
  {1961}{\natexlab{a}})}\BibitemShut {NoStop}%
\bibitem [{\citenamefont {Nambu}\ and\ \citenamefont
  {Jona-Lasinio}(1961{\natexlab{b}})}]{Nam612}%
  \BibitemOpen
  \bibfield  {author} {\bibinfo {author} {\bibfnamefont {Y.}~\bibnamefont
  {Nambu}}\ and\ \bibinfo {author} {\bibfnamefont {G.}~\bibnamefont
  {Jona-Lasinio}},\ }\href {\doibase 10.1103/PhysRev.124.246} {\bibfield
  {journal} {\bibinfo  {journal} {Phys. Rev.}\ }\textbf {\bibinfo {volume}
  {124}},\ \bibinfo {pages} {246} (\bibinfo {year}
  {1961}{\natexlab{b}})}\BibitemShut {NoStop}%
\bibitem [{\citenamefont {Asakawa}\ and\ \citenamefont {Yazaki}(1989)}]{Asa89}%
  \BibitemOpen
  \bibfield  {author} {\bibinfo {author} {\bibfnamefont {M.}~\bibnamefont
  {Asakawa}}\ and\ \bibinfo {author} {\bibfnamefont {K.}~\bibnamefont
  {Yazaki}},\ }\href {\doibase 10.1016/0375-9474(89)90002-X} {\bibfield
  {journal} {\bibinfo  {journal} {Nucl. Phys.}\ }\textbf {\bibinfo {volume}
  {A504}},\ \bibinfo {pages} {668} (\bibinfo {year} {1989})}\BibitemShut
  {NoStop}%
\bibitem [{\citenamefont {Carignano}\ \emph {et~al.}(2010)\citenamefont
  {Carignano}, \citenamefont {Nickel},\ and\ \citenamefont {Buballa}}]{Car10}%
  \BibitemOpen
  \bibfield  {author} {\bibinfo {author} {\bibfnamefont {S.}~\bibnamefont
  {Carignano}}, \bibinfo {author} {\bibfnamefont {D.}~\bibnamefont {Nickel}}, \
  and\ \bibinfo {author} {\bibfnamefont {M.}~\bibnamefont {Buballa}},\ }\href
  {\doibase 10.1103/PhysRevD.82.054009} {\bibfield  {journal} {\bibinfo
  {journal} {Phys. Rev.}\ }\textbf {\bibinfo {volume} {D82}},\ \bibinfo {pages}
  {054009} (\bibinfo {year} {2010})},\ \Eprint {http://arxiv.org/abs/1007.1397}
  {arXiv:1007.1397 [hep-ph]} \BibitemShut {NoStop}%
\bibitem [{\citenamefont {Bratovic}\ \emph {et~al.}(2013)\citenamefont
  {Bratovic}, \citenamefont {Hatsuda},\ and\ \citenamefont {Weise}}]{Bra12}%
  \BibitemOpen
  \bibfield  {author} {\bibinfo {author} {\bibfnamefont {N.~M.}\ \bibnamefont
  {Bratovic}}, \bibinfo {author} {\bibfnamefont {T.}~\bibnamefont {Hatsuda}}, \
  and\ \bibinfo {author} {\bibfnamefont {W.}~\bibnamefont {Weise}},\ }\href
  {\doibase 10.1016/j.physletb.2013.01.003} {\bibfield  {journal} {\bibinfo
  {journal} {Phys. Lett.}\ }\textbf {\bibinfo {volume} {B719}},\ \bibinfo
  {pages} {131} (\bibinfo {year} {2013})},\ \Eprint
  {http://arxiv.org/abs/1204.3788} {arXiv:1204.3788 [hep-ph]} \BibitemShut
  {NoStop}%
\bibitem [{\citenamefont {Stephanov}(2004)}]{Stephanov:2004wx}%
  \BibitemOpen
  \bibfield  {author} {\bibinfo {author} {\bibfnamefont {M.~A.}\ \bibnamefont
  {Stephanov}},\ }\bibfield  {booktitle} {\emph {\bibinfo {booktitle}
  {{Non-perturbative quantum chromodynamics. Proceedings, 8th Workshop, Paris,
  France, June 7-11, 2004}}},\ }\href {\doibase 10.1142/S0217751X05027965}
  {\bibfield  {journal} {\bibinfo  {journal} {Prog. Theor. Phys. Suppl.}\
  }\textbf {\bibinfo {volume} {153}},\ \bibinfo {pages} {139} (\bibinfo {year}
  {2004})},\ \bibinfo {note} {[Int. J. Mod. Phys.A20,4387(2005)]},\ \Eprint
  {http://arxiv.org/abs/hep-ph/0402115} {arXiv:hep-ph/0402115 [hep-ph]}
  \BibitemShut {NoStop}%
\bibitem [{\citenamefont {Stephanov}(2006)}]{Stephanov:2007fk}%
  \BibitemOpen
  \bibfield  {author} {\bibinfo {author} {\bibfnamefont {M.~A.}\ \bibnamefont
  {Stephanov}},\ }\bibfield  {booktitle} {\emph {\bibinfo {booktitle}
  {{Proceedings, 24th International Symposium on Lattice Field Theory (Lattice
  2006): Tucson, USA, July 23-28, 2006}}},\ }\href {\doibase
  10.22323/1.032.0024} {\bibfield  {journal} {\bibinfo  {journal} {PoS}\
  }\textbf {\bibinfo {volume} {LAT2006}},\ \bibinfo {pages} {024} (\bibinfo
  {year} {2006})},\ \Eprint {http://arxiv.org/abs/hep-lat/0701002}
  {arXiv:hep-lat/0701002 [hep-lat]} \BibitemShut {NoStop}%
\bibitem [{\citenamefont {Fukushima}(2008)}]{Fukushima:2008wg}%
  \BibitemOpen
  \bibfield  {author} {\bibinfo {author} {\bibfnamefont {K.}~\bibnamefont
  {Fukushima}},\ }\href {\doibase 10.1103/PhysRevD.77.114028,
  10.1103/PhysRevD.78.039902} {\bibfield  {journal} {\bibinfo  {journal} {Phys.
  Rev.}\ }\textbf {\bibinfo {volume} {D77}},\ \bibinfo {pages} {114028}
  (\bibinfo {year} {2008})},\ \bibinfo {note} {[Erratum: Phys.
  Rev.D78,039902(2008)]},\ \Eprint {http://arxiv.org/abs/0803.3318}
  {arXiv:0803.3318 [hep-ph]} \BibitemShut {NoStop}%
\bibitem [{\citenamefont {Fukushima}\ and\ \citenamefont
  {Sasaki}(2013)}]{Fukushima:2013rx}%
  \BibitemOpen
  \bibfield  {author} {\bibinfo {author} {\bibfnamefont {K.}~\bibnamefont
  {Fukushima}}\ and\ \bibinfo {author} {\bibfnamefont {C.}~\bibnamefont
  {Sasaki}},\ }\href {\doibase 10.1016/j.ppnp.2013.05.003} {\bibfield
  {journal} {\bibinfo  {journal} {Prog. Part. Nucl. Phys.}\ }\textbf {\bibinfo
  {volume} {72}},\ \bibinfo {pages} {99} (\bibinfo {year} {2013})},\ \Eprint
  {http://arxiv.org/abs/1301.6377} {arXiv:1301.6377 [hep-ph]} \BibitemShut
  {NoStop}%
\bibitem [{\citenamefont {Baym}\ \emph {et~al.}(2018)\citenamefont {Baym},
  \citenamefont {Hatsuda}, \citenamefont {Kojo}, \citenamefont {Powell},
  \citenamefont {Song},\ and\ \citenamefont {Takatsuka}}]{Baym:2017whm}%
  \BibitemOpen
  \bibfield  {author} {\bibinfo {author} {\bibfnamefont {G.}~\bibnamefont
  {Baym}}, \bibinfo {author} {\bibfnamefont {T.}~\bibnamefont {Hatsuda}},
  \bibinfo {author} {\bibfnamefont {T.}~\bibnamefont {Kojo}}, \bibinfo {author}
  {\bibfnamefont {P.~D.}\ \bibnamefont {Powell}}, \bibinfo {author}
  {\bibfnamefont {Y.}~\bibnamefont {Song}}, \ and\ \bibinfo {author}
  {\bibfnamefont {T.}~\bibnamefont {Takatsuka}},\ }\href {\doibase
  10.1088/1361-6633/aaae14} {\bibfield  {journal} {\bibinfo  {journal} {Rept.
  Prog. Phys.}\ }\textbf {\bibinfo {volume} {81}},\ \bibinfo {pages} {056902}
  (\bibinfo {year} {2018})},\ \Eprint {http://arxiv.org/abs/1707.04966}
  {arXiv:1707.04966 [astro-ph.HE]} \BibitemShut {NoStop}%
\bibitem [{\citenamefont {Sun}\ \emph {et~al.}(2020{\natexlab{a}})\citenamefont
  {Sun}, \citenamefont {Ko}, \citenamefont {Cao},\ and\ \citenamefont
  {Li}}]{Sun:2020bbn}%
  \BibitemOpen
  \bibfield  {author} {\bibinfo {author} {\bibfnamefont {K.-J.}\ \bibnamefont
  {Sun}}, \bibinfo {author} {\bibfnamefont {C.-M.}\ \bibnamefont {Ko}},
  \bibinfo {author} {\bibfnamefont {S.}~\bibnamefont {Cao}}, \ and\ \bibinfo
  {author} {\bibfnamefont {F.}~\bibnamefont {Li}},\ }\href@noop {} {\
  (\bibinfo {year} {2020}{\natexlab{a}})},\ \Eprint
  {http://arxiv.org/abs/2004.05754} {arXiv:2004.05754 [nucl-th]} \BibitemShut
  {NoStop}%
\bibitem [{\citenamefont {Rischke}\ \emph {et~al.}(1995)\citenamefont
  {Rischke}, \citenamefont {Pursun}, \citenamefont {Maruhn}, \citenamefont
  {Stoecker},\ and\ \citenamefont {Greiner}}]{Rischke:1995pe}%
  \BibitemOpen
  \bibfield  {author} {\bibinfo {author} {\bibfnamefont {D.~H.}\ \bibnamefont
  {Rischke}}, \bibinfo {author} {\bibfnamefont {Y.}~\bibnamefont {Pursun}},
  \bibinfo {author} {\bibfnamefont {J.~A.}\ \bibnamefont {Maruhn}}, \bibinfo
  {author} {\bibfnamefont {H.}~\bibnamefont {Stoecker}}, \ and\ \bibinfo
  {author} {\bibfnamefont {W.}~\bibnamefont {Greiner}},\ }\href@noop {}
  {\bibfield  {journal} {\bibinfo  {journal} {Acta Phys. Hung. A}\ }\textbf
  {\bibinfo {volume} {1}},\ \bibinfo {pages} {309} (\bibinfo {year} {1995})},\
  \Eprint {http://arxiv.org/abs/nucl-th/9505014} {arXiv:nucl-th/9505014}
  \BibitemShut {NoStop}%
\bibitem [{\citenamefont {Stoecker}(2005)}]{Stoecker:2004qu}%
  \BibitemOpen
  \bibfield  {author} {\bibinfo {author} {\bibfnamefont {H.}~\bibnamefont
  {Stoecker}},\ }\href {\doibase 10.1016/j.nuclphysa.2004.12.074} {\bibfield
  {journal} {\bibinfo  {journal} {Nucl. Phys. A}\ }\textbf {\bibinfo {volume}
  {750}},\ \bibinfo {pages} {121} (\bibinfo {year} {2005})},\ \Eprint
  {http://arxiv.org/abs/nucl-th/0406018} {arXiv:nucl-th/0406018} \BibitemShut
  {NoStop}%
\bibitem [{\citenamefont {Sun}\ \emph {et~al.}(2017)\citenamefont {Sun},
  \citenamefont {Chen}, \citenamefont {Ko},\ and\ \citenamefont
  {Xu}}]{Sun:2017xrx}%
  \BibitemOpen
  \bibfield  {author} {\bibinfo {author} {\bibfnamefont {K.-J.}\ \bibnamefont
  {Sun}}, \bibinfo {author} {\bibfnamefont {L.-W.}\ \bibnamefont {Chen}},
  \bibinfo {author} {\bibfnamefont {C.~M.}\ \bibnamefont {Ko}}, \ and\ \bibinfo
  {author} {\bibfnamefont {Z.}~\bibnamefont {Xu}},\ }\href {\doibase
  10.1016/j.physletb.2017.09.056} {\bibfield  {journal} {\bibinfo  {journal}
  {Phys. Lett.}\ }\textbf {\bibinfo {volume} {B774}},\ \bibinfo {pages} {103}
  (\bibinfo {year} {2017})},\ \Eprint {http://arxiv.org/abs/1702.07620}
  {arXiv:1702.07620 [nucl-th]} \BibitemShut {NoStop}%
\bibitem [{\citenamefont {Sun}\ \emph {et~al.}(2018)\citenamefont {Sun},
  \citenamefont {Chen}, \citenamefont {Ko}, \citenamefont {Pu},\ and\
  \citenamefont {Xu}}]{Sun:2018jhg}%
  \BibitemOpen
  \bibfield  {author} {\bibinfo {author} {\bibfnamefont {K.-J.}\ \bibnamefont
  {Sun}}, \bibinfo {author} {\bibfnamefont {L.-W.}\ \bibnamefont {Chen}},
  \bibinfo {author} {\bibfnamefont {C.~M.}\ \bibnamefont {Ko}}, \bibinfo
  {author} {\bibfnamefont {J.}~\bibnamefont {Pu}}, \ and\ \bibinfo {author}
  {\bibfnamefont {Z.}~\bibnamefont {Xu}},\ }\href {\doibase
  10.1016/j.physletb.2018.04.035} {\bibfield  {journal} {\bibinfo  {journal}
  {Phys. Lett.}\ }\textbf {\bibinfo {volume} {B781}},\ \bibinfo {pages} {499}
  (\bibinfo {year} {2018})},\ \Eprint {http://arxiv.org/abs/1801.09382}
  {arXiv:1801.09382 [nucl-th]} \BibitemShut {NoStop}%
\bibitem [{\citenamefont {Sun}\ \emph {et~al.}(2020{\natexlab{b}})\citenamefont
  {Sun}, \citenamefont {Ko}, \citenamefont {Li}, \citenamefont {Xu},\ and\
  \citenamefont {Chen}}]{Sun:2020pjz}%
  \BibitemOpen
  \bibfield  {author} {\bibinfo {author} {\bibfnamefont {K.-J.}\ \bibnamefont
  {Sun}}, \bibinfo {author} {\bibfnamefont {C.~M.}\ \bibnamefont {Ko}},
  \bibinfo {author} {\bibfnamefont {F.}~\bibnamefont {Li}}, \bibinfo {author}
  {\bibfnamefont {J.}~\bibnamefont {Xu}}, \ and\ \bibinfo {author}
  {\bibfnamefont {L.-W.}\ \bibnamefont {Chen}},\ }\href@noop {} {\  (\bibinfo
  {year} {2020}{\natexlab{b}})},\ \Eprint {http://arxiv.org/abs/2006.08929}
  {arXiv:2006.08929 [nucl-th]} \BibitemShut {NoStop}%
\bibitem [{\citenamefont {Li}\ and\ \citenamefont {Ko}(2017)}]{Li:2016uvu}%
  \BibitemOpen
  \bibfield  {author} {\bibinfo {author} {\bibfnamefont {F.}~\bibnamefont
  {Li}}\ and\ \bibinfo {author} {\bibfnamefont {C.~M.}\ \bibnamefont {Ko}},\
  }\href {\doibase 10.1103/PhysRevC.95.055203} {\bibfield  {journal} {\bibinfo
  {journal} {Phys. Rev. C}\ }\textbf {\bibinfo {volume} {95}},\ \bibinfo
  {pages} {055203} (\bibinfo {year} {2017})},\ \Eprint
  {http://arxiv.org/abs/1606.05012} {arXiv:1606.05012 [nucl-th]} \BibitemShut
  {NoStop}%
\bibitem [{\citenamefont {Bzdak}\ \emph {et~al.}(2020)\citenamefont {Bzdak},
  \citenamefont {Esumi}, \citenamefont {Koch}, \citenamefont {Liao},
  \citenamefont {Stephanov},\ and\ \citenamefont {Xu}}]{Bzdak:2019pkr}%
  \BibitemOpen
  \bibfield  {author} {\bibinfo {author} {\bibfnamefont {A.}~\bibnamefont
  {Bzdak}}, \bibinfo {author} {\bibfnamefont {S.}~\bibnamefont {Esumi}},
  \bibinfo {author} {\bibfnamefont {V.}~\bibnamefont {Koch}}, \bibinfo {author}
  {\bibfnamefont {J.}~\bibnamefont {Liao}}, \bibinfo {author} {\bibfnamefont
  {M.}~\bibnamefont {Stephanov}}, \ and\ \bibinfo {author} {\bibfnamefont
  {N.}~\bibnamefont {Xu}},\ }\href {\doibase 10.1016/j.physrep.2020.01.005}
  {\bibfield  {journal} {\bibinfo  {journal} {Phys. Rept.}\ }\textbf {\bibinfo
  {volume} {853}},\ \bibinfo {pages} {1} (\bibinfo {year} {2020})},\ \Eprint
  {http://arxiv.org/abs/1906.00936} {arXiv:1906.00936 [nucl-th]} \BibitemShut
  {NoStop}%
\bibitem [{\citenamefont {Steinheimer}\ and\ \citenamefont
  {Randrup}(2012)}]{Steinheimer:2012gc}%
  \BibitemOpen
  \bibfield  {author} {\bibinfo {author} {\bibfnamefont {J.}~\bibnamefont
  {Steinheimer}}\ and\ \bibinfo {author} {\bibfnamefont {J.}~\bibnamefont
  {Randrup}},\ }\href {\doibase 10.1103/PhysRevLett.109.212301} {\bibfield
  {journal} {\bibinfo  {journal} {Phys. Rev. Lett.}\ }\textbf {\bibinfo
  {volume} {109}},\ \bibinfo {pages} {212301} (\bibinfo {year} {2012})},\
  \Eprint {http://arxiv.org/abs/1209.2462} {arXiv:1209.2462 [nucl-th]}
  \BibitemShut {NoStop}%
\bibitem [{\citenamefont {Adamczyk}\ and\ \citenamefont
  {etc.}(2017)}]{PhysRevC.96.044904}%
  \BibitemOpen
  \bibfield  {author} {\bibinfo {author} {\bibfnamefont {L.}~\bibnamefont
  {Adamczyk}}\ and\ \bibinfo {author} {\bibnamefont {etc.}} (\bibinfo
  {collaboration} {STAR Collaboration}),\ }\href {\doibase
  10.1103/PhysRevC.96.044904} {\bibfield  {journal} {\bibinfo  {journal} {Phys.
  Rev. C}\ }\textbf {\bibinfo {volume} {96}},\ \bibinfo {pages} {044904}
  (\bibinfo {year} {2017})}\BibitemShut {NoStop}%
\end{thebibliography}%
\end{document}